%% file: aaaa.tex
\documentclass[10pt,english,prd,aps,superscriptaddress,twocolumn,showpacs,footinbib]{revtex4-1}

\usepackage{silence}
\usepackage[utf8]{inputenc}
\usepackage[T2A]{fontenc}
\usepackage{graphicx} 
\usepackage{dcolumn}  
\usepackage{bm}       
\usepackage{amssymb}   
\usepackage{amsthm}
\usepackage[english]{babel}
\usepackage{float}
\usepackage{verbatim}
\usepackage{amsmath}
\usepackage{setspace}
\usepackage{epstopdf}
\graphicspath{{figures/}}
\usepackage{siunitx}
\usepackage{hyperref}
\usepackage{mathtools}
\usepackage{nicefrac, xfrac}

\DeclarePairedDelimiter{\ceil}{\lceil}{\rceil}
\DeclarePairedDelimiter{\floor}{\lfloor}{\rfloor}
\DeclareMathOperator{\Tr}{Tr}

\DeclareMathOperator{\me}{e}

\newcommand{\defeq}{\vcentcolon=}

\DeclarePairedDelimiter\bra{\langle}{\rvert}
\DeclarePairedDelimiter\ket{\lvert}{\rangle}
\DeclarePairedDelimiterX\braket[2]{\langle}{\rangle}{#1 \delimsize\vert #2}

\graphicspath{ {./image/} }
\usepackage{subfiles} 

\newcommand{\sll}{\sum_{\begin{subarray}{l} \rho \in \mathcal{L}\\ \rho \ge r \end{subarray}}}

\theoremstyle{definition}
\newtheorem{theorem}{Theorem}
\newtheorem*{remark}{Remark}
\newtheorem{proposition}{Proposition}
\newtheorem*{proposition*}{Proposition}

\makeatletter
\newcommand{\proofpart}[2]{%
  \par
  \addvspace{\medskipamount}%
  \noindent\emph{Part #1: #2}\par\nobreak
  \addvspace{\smallskipamount}%
  \@afterheading
}

\makeatother

\usepackage{natbib} 

\begin{document}

\title{Dephasing in the central spin problem with long-range Ising spin-bath coupling}

\author{K. I. O. Ben Attar} 
\affiliation{Institute  of Applied Physics, The Hebrew University of Jerusalem, Jerusalem 9190401, Israel}

\author{N. Bar-Gill}
\affiliation{Institute  of Applied Physics, The Hebrew University of Jerusalem, Jerusalem 9190401, Israel}
\affiliation{Racah Institute of Physics, The Hebrew University of Jerusalem, Jerusalem 9190401, Israel}

\date{\today}

\subfile{./abstract}

\maketitle

\subfile{./introduction}

\section{Mathematical preliminary}

\subfile{./system}

\subfile{./deloneSumIntegralBound}

\section{Ramsey dephasing for inverse power-law coupling}

\subfile{./globalBound}

\subfile{"./counterExample"}

\section{Conclusions}

\subfile{"./conclusion"}

\section{Acknowledgements}
N.B. acknowledges financial support by the European Commission’s Horizon Europe Framework Programme under the Research and Innovation Actions GA No. 828946–PATHOS and GA No.~101070546–MUQUABIS. N.B.~also acknowledges financial support by the Carl Zeiss Stiftung (HYMMS wildcard), the Ministry of Science and Technology, Israel, the innovation authority (Project No.~70033), and the ISF (Grants No.~1380/21 and No.~3597/21).

\bibliography{references}

\subfile{./anexe/anexe}

\end{document}

%% file: abstract.tex
\begin{abstract}
The study of coherence dynamics in open quantum systems, specifically addressing various physical realizations of quantum systems and environments, is a long-standing and central pillar of quantum science and technology. As such, a large body of work establishes a firm theoretical understanding of these processes. Nevertheless, a fundamental aspect of decoherence dynamics, namely the central limit theorem of qubit dephasing in the central spin model, which leads to a Gaussian approximation, lacks formal proof in realistically relevant scenarios. Here we prove this approximation for a bath depicted by an Ising spin system, in the presence of disorder and several (most relevant) functional forms of qubit-bath coupling. Importantly, we show that in certain cases, namely for short-range (exponentially decaying) coupling, this approximation breaks. These results provide a theoretical framework for studying decoherence dynamics in various systems and lead to insights into dephasing behavior with implications for applications in quantum information, quantum computing, and other quantum technologies.
\end{abstract}

%% file: introduction.tex
\section{Introduction}

The dynamics of open quantum systems, and specifically dephasing dynamics of a quantum spin (qubit) coupled to a spin bath, are a central pillar of quantum science and technology \cite{Nielsen_Chuang_2010, Breuer2002-ik, Gaudin1976, PhysRevB.76.014304, PhysRevA.72.052113, PhysRevB.86.035452}. As such, these processes have been studied extensively both theoretically and experimentally, and standard schemes are used to characterize them in realistic scenarios. The most common example is Ramsey Spectroscopy \cite{ramsey} (sometimes referred to as free induction decay, or FID, mostly in the context of NMR and spin systems), which enables a direct measurement of the dephasing time of a two-level system. For example, this is an essential step in characterizing qubits in quantum computers (such as those developed by IBM, Google, etc.) \cite{Penshin2024, PRXQuantum.3.020357, Ladd2010,PhysRevB.72.134519}. 

As an important model system, which is highly relevant for a broad range of realistic scenarios, we address the central spin model, in which a quantum spin interacts with a spin bath.
In this context, gaining insights into the behavior of a central spin when it interacts with an infinite spin bath becomes imperative to grasp the intricacies of such experiments fully. 

Considering each spin as an uncorrelated random variable, one might initially assume that the spin bath could be approximated by a Gaussian random variable. In most cases, this is the accepted assumption \cite{deSousa2009}. However, this naive perspective falls short due to the dependence of the coupling on the distance between spins. 
Moreover, in cases where the system is not infinite, the dynamics follow the theory of quasi-periodic functions, causing the system to return arbitrarily close to its initial state over time. Consequently, the dynamics of the system never uniformly approach a Gaussian distribution.

Therefore, here we address the central spin problem in a mathematically rigorous way, encompassing disorder, different dimensionalities, and various forms of interactions (with emphasis on inverse power law couplings).
We demonstrate that in the case of an infinite spin system, if the couplings are, to a certain extent, "approximately" equal, the Gaussian approximation holds well. By "approximately" equal, we refer to interactions that are long-range enough, allowing each spin to contribute equally to the dynamics of the central spin. 
This bound is specifically applicable to long-range interactions and does not hold for exponentially decaying interactions, where the contribution of closer spins becomes dominant and significantly affects the dynamics. 

The paper is structured as follows: 
First, we describe the system we analyze, the terms and symbols we employ, and the basic assumptions and techniques.
Then we prove an estimate for the effect of the bath (disordered spins) on the central spin, through the sum on a Delone set of $\mathbb{R}^d$.
Based on this, we show the convergence of the dephasing to Gaussian behavior of the dynamic profile, under certain assumptions.
Finally, we provide counter-examples (exponentially decaying interaction) for which Gaussian dephasing dynamics are not a good approximation.

%% file: system.tex
\subsection{Description of the system}

\begin{figure}
    \centering
    \includegraphics[width=0.5\textwidth]{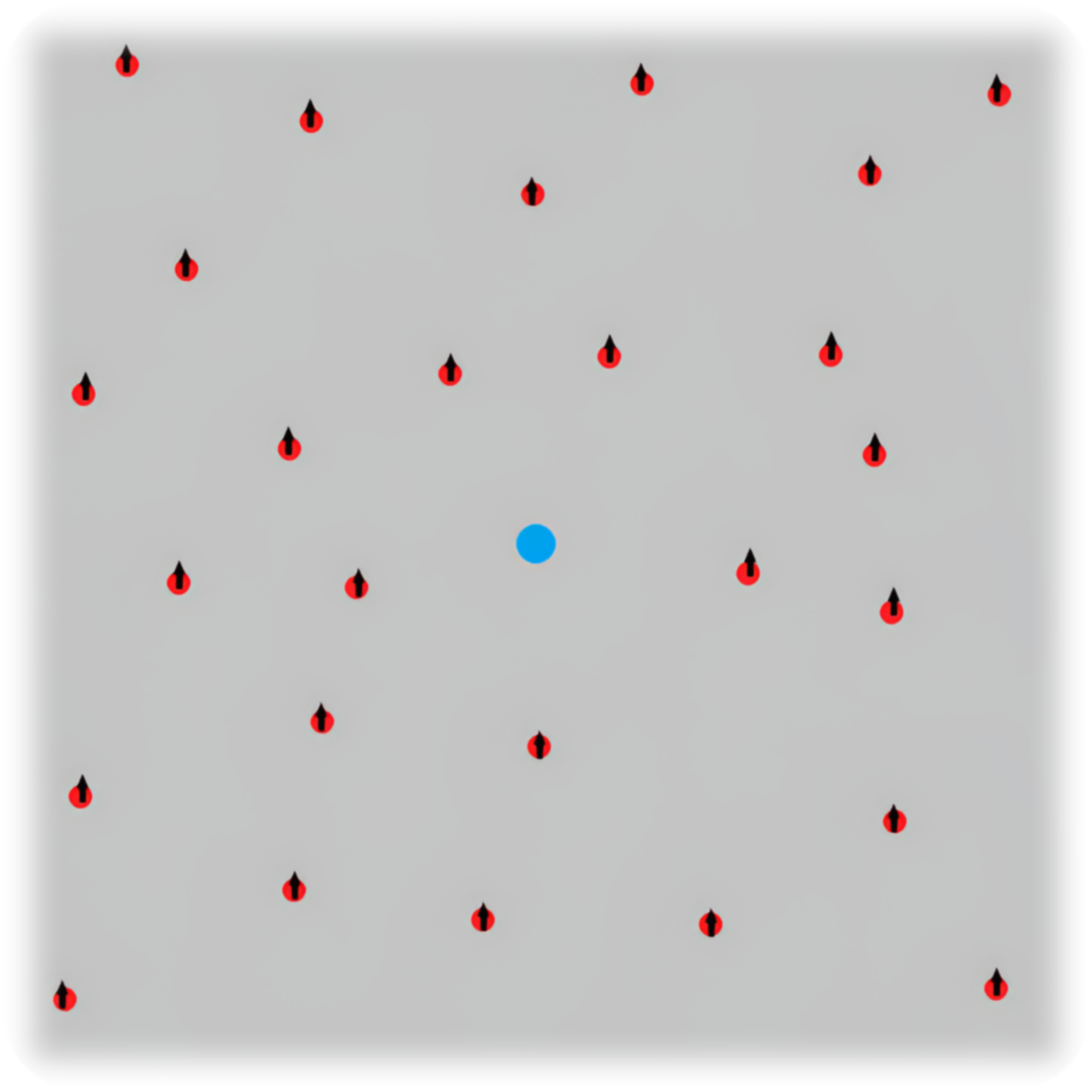}
    \caption{The system considered consists of a central spin (represented by the blue big dot) surrounded by an infinite number of spins (represented by orange small dots)  consisting of the spin bath. In this example all spins are polarized in one direction.}
    \label{fig:systemSchem}
\end{figure}

In this article, we consider a physical system comprising of a central spin $\frac{1}{2}$ situated at the origin (Fig. \ref{fig:systemSchem}). The central spin is coupled to an infinite number of spin $\frac{1}{2}$ particles which constitute the bath. The position of each spin in the bath belongs to $\mathcal{L}$, a subset of $\mathbb{R}^d$. The strength of the coupling between the central spin and each spin in the bath will depend uniquely on its distance $\rho$ to the central spin through a function $A(\rho)$.

We will suppose in general that the couplings are square summable, that is $\sum_{\rho \in \mathcal{L}}A(\rho)^2 < +\infty$. The Hamiltonian will be:

\[
H = S_z\sum_{\rho \in \mathcal{L}}A(\rho) I_z^{(\rho)}
\]

\begin{remark}
    For the case of summable couplings, there is no ambiguity for the sum as it converges in the norm topology. For couplings that are not summable but square-summable, the sum is still meaningful in the strong resolvent sense. See the Annex \ref{convOfSum} for further details.
\end{remark}

From a mathematical perspective of many-body systems, a common description employs a $C^*$-algebra rather than a Hilbert space; this point of view, introduced by von Neumann, is sometimes also called algebraic quantum mechanics \cite{vNeumann1930}.

A $C^*$-algebra is a normed associative algebra that is also a complete metric space and possesses an involutive antiautomorphism $*$, such that for every complex number $\lambda$ and element $a$ of the algebra, we have $(\lambda a)^* = \overline{\lambda} a^*$ and $||x^* x|| = ||x^*|| \, ||x||$. The latter equality is called the $C^*$-identity. An example of a $C^*$-algebra is the set of bounded operators on a Hilbert space, which is more commonly used in quantum physics \cite{pieter}.

We do not intend to provide here an extensive introduction to $C^*$-algebras for the sake of conciseness, yet the presentation is self-sufficient. For a brief overview of the $C^*$-algebras pertaining to infinite-spin systems, we recommend consulting the reference \cite{pieter}. For a more comprehensive understanding, consider exploring the references \cite{blackadar} and \cite{bratAndRobin}.

In this framework, observables of the system are elements of a $C^*$-algebra $\mathcal{U}$ and the set of the states is the set of positive linear functional of norm 1 on this algebra. Positivity means that for all $a\in\mathcal{U}$ we have $\rho(a^* a)\ge0$ and for $C^*$-algebra with an identity the normalization amount to the equality $\rho(1)=1$.
From a state $\rho$, we can associate a Hilbert space $\mathcal{H}$, an $*$-isomorphism $\pi$ from the $C^*$-algebra to the set of bounded operators of $\mathcal{H}$ and a vector $\ket{\Omega}$ in $\mathcal{H}$, such that $\rho(A)=\bra{\Omega}\pi(A)\ket{\Omega}$ and that the set $\{\pi(a)\ket{\Omega}|a\in\mathcal{U}\}$ is dense in $\mathcal{U}$. Such a triplet $(\mathcal{H},\pi,\ket{\Omega})$ is unique up to a unitary equivalence and is called the GNS (Gelfand–Naimark–Segal) construction for the state $\rho$.

For each bounded open $\mathcal{O}$ subset of $\mathbb{R}^d$ consider the $C^*$-algebra $\mathcal{U}_\mathcal{O}$ consisting of spins positioned inside $\mathcal{O}$. This $C^*$-algebra is $*$-isomorphic to the $C^*$-algebra of bounded operators of a separable Hilbert space of dimension $2^n$, where $n$ is the number of spins inside $\mathcal{O}$. If we have the relation $\mathcal{O}\subset\mathcal{V}$ we also have $\mathcal{U}_\mathcal{O}\subset\mathcal{U}_\mathcal{V}$. Now let $\mathcal{U}_\mathrm{loc}$ be the set of observable $x$ such that there is an open set such that $x\in\mathcal{U}_\mathcal{O}$. The $C^*$-algebra $\mathcal{U}$ will be defined as the closure of $\mathcal{U}_\mathrm{loc}$.

Here we assume that the initial state of the bath is the totally unpolarized state. The construction of the GNS representation of this state is described in Annex \ref{an:GNS}. 

The set of states $\ket{\alpha}$ is a Hilbert basis of this representation (see Annex for details) and this representation can be thought of as the set of states where only a finite number of spin are not unpolarized. As such in the rest of the article, we consider the different observables of the system as operators in the Hilbert space $\mathcal{H}$. We will denote $\Vec{S}$ the Pauli vector of the central spin and if we note the position of a spin of the bath as $\rho$ (with respect to the central spin), we will denote the representation of the Pauli vector in the Hilbert space $\mathcal{H}$ as $\Vec{I}^{(\rho)}$.

\subsection{Dynamical characterization}

An interesting quantity that can be measured in real experiments is the Ramsey profile, which is the dephasing or decoherence profile of the central spin \cite{ramsey}.
Given the central spin fully polarized in the $z$-direction and the spin bath in a state $\sigma$ one can perform an initial and final pulse to compute the $S_x$ and $S_y$ dynamic profile, obtaining the quantity $S_+ = S_x + iS_y$:

\[
\sigma\left(\me^{iHt} S_+ \me^{-iHt}\right)= \sigma\left(\me^{it\sum_{\rho \in \mathcal{L}}A(\rho) I_z^{(\rho)}}\right)
\]

Note that every state and operator is associated with a probability distribution $P$ such that for every Borelian function $\bra{\psi}f(A)\ket{\psi}=\int f(x)\mathrm{d}P(x)$ \cite{theBible}.
Thus, the previous quantity can be interpreted as the characteristic function of the probability measure $P$, and therefore analyzing the Ramsey profile of the central spin allows us to extract information about the spin bath.

In order to compute the measurable dephasing dynamics of the central spin, we study the limit of the normalized sum of all coupling operators of the bath spins to the central spin:

\[B_r = \sum_{\begin{subarray}{l} \rho \in \mathcal{L}\\ \rho \ge r \end{subarray}}  \frac{A(\rho) I_z^{(\rho)}}{\sqrt{\sum_{\begin{subarray}{l} \rho \in \mathcal{L}\\ \rho \ge r \end{subarray}} A(\rho)^2}}\]

Taking initial and final states $\ket{\alpha}$ and $\ket{\beta}$, as defined before, there is an $R$ such that all spins with $r>R$ are unpolarized and for such $r$:

\[
\bra{\alpha}\me^{iB_r t}\ket{\beta} = \prod_{\begin{subarray}{l} \rho \in \mathcal{L}\\ \rho \ge r \end{subarray}} \cos\left(\frac{A(\rho) t}{\sqrt{\sum_{\begin{subarray}{l} \rho \in \mathcal{L}\\ \rho \ge r \end{subarray}} A(\rho)^2}}\right)
\]

Therefore we can define the function corresponding to the measurable Ramsey dephasing:

\[
C_r(t) = \prod_{\begin{subarray}{l} \rho \in \mathcal{L}\\ \rho \ge r \end{subarray}} \cos\left(\frac{A(\rho) t}{\sqrt{\sum_{\begin{subarray}{l} \rho \in \mathcal{L}\\ \rho \ge r \end{subarray}} A(\rho)^2}}\right)
\]

\textit{The function $C_r$ is the main quantity that will be studied here, and in particular its asymptotic behavior as $r$ goes to infinity}.

\subsection{Spatial configuration of the spins and the Delone set}

\begin{figure}[tbh]
    \centering
    \includegraphics[width=0.5\textwidth]{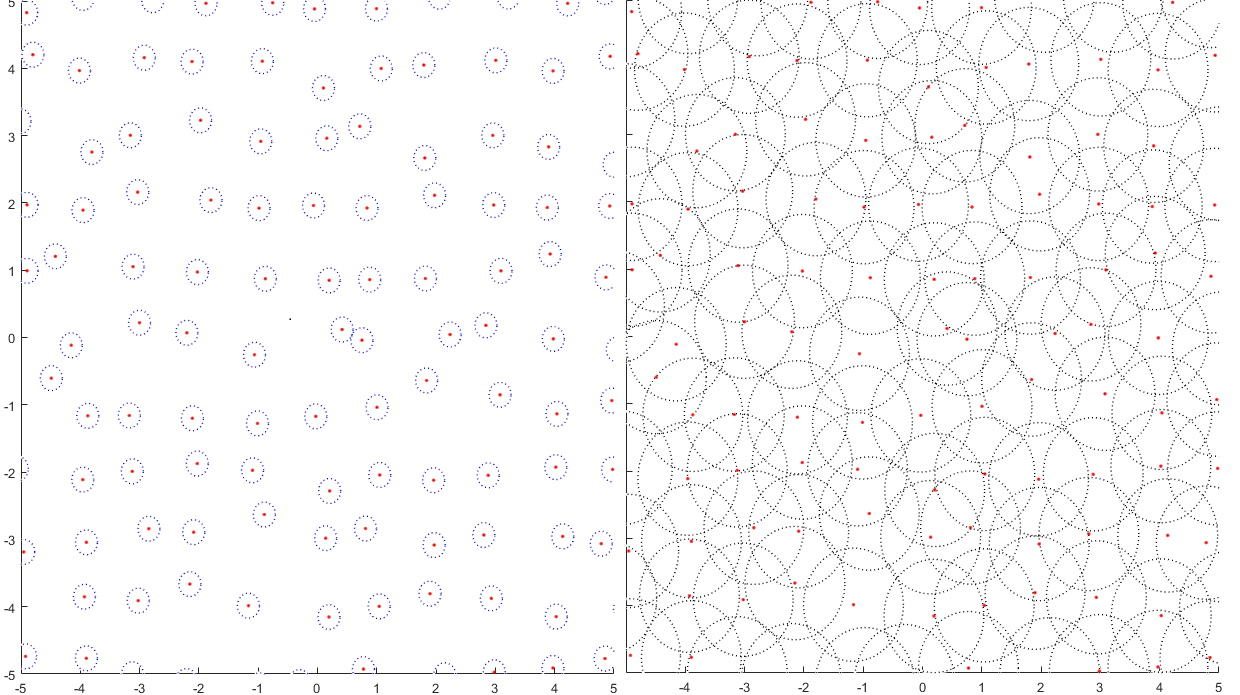}
    \caption{Illustration of the concept of packing radius and covering radius for a Delone set. The label of the axis corresponds to the spatial position of the spin in arbitrary units the central spin being situated at $(0,0)$. In each figure, the red dot corresponds to a spin of the spin bath. In the right picture is illustrated the packing radius where you can see that no balls intersect each other. In the left picture is illustrated the covering radius which is the smallest radius such that all the space is covered by the balls.}
    \label{fig:delone}
\end{figure}

In considering the spatial distribution of the bath spins we employ a Delone construction.
One of the mathematical definitions of a homogeneously distributed set of particles in a metric space $X$ could be a Delone space \cite{delaunay1934sphere, Ito2015}, see Fig. \ref{fig:delone}. Importantly, this construction allows for a rigorous consideration of varied spatial distributions of the bath spins, incorporating effects of disorder (in certain limits as detailed below).

Let's note $B(r,\epsilon)$ the open ball of center $r$ and radius $\epsilon$ and $\Bar{B}(r,\epsilon)$ the closed one.
A space $\mathcal{L}$ is called a Delone set of $X$ if there exist two positive real numbers $\hat{r}$ and $\check{r}$ such that for all $r\in X$ we have:

\[
\left|
\mathcal{L}
\cap
B(r,\check{r})
\right| \le 1 \text{ and }
\left|
\mathcal{L}
\cap
\Bar{B}(r,\hat{r})
\right| \ge 1 .
\]

Intuitively we can interpret $2\check{r}$ as a lower bound for the distance between two spins. Indeed, if two sites have a distance between them smaller than $2\check{r}$, then if we take a ball of radius $2\check{r}$ centered at the midpoint between these two sites, this ball will have two sites inside it which leads to a contradiction. This ensures that the spin concentration does not diverge at some point in the space.

On the other hand, we can interpret $\hat{r}$ as an upper bound for the size of a hole in the configuration, and since the size of such holes is bounded, the spins are well scattered in the space.

In the following sections, we will study the function $C_r$ where $\mathcal{L}$ is a Delone set of $\mathbb{R}^d$.

%% file: deloneSumIntegralBound.tex
\subsection{Delone Sum-Integral Bound}

Here, we aim to establish the asymptotic equivalence between the summation over a Delone set and the corresponding integral across real space. More precisely we have:

\begin{theorem}
Let $\mathcal{L}$ be a Delone set with a covering radius $\check{r}$ and packing radius $\hat{r}$.
Assume $f$ is a function defined over positive real numbers, exhibiting positive values, and satisfying the condition that the mapping $r \mapsto r^{d-1}f(r)$ is decreasing. Then, for $r\ge 3\hat{r}$ we have:

\begin{equation*}
\boxed{
\frac{d}{3^d \hat{r}^d}\displaystyle\int\limits_{r+\hat{r}}^{+\infty} \rho^{d-1}f(\rho) \,\mathrm{d}\rho  \le \sll f(\rho) \le \frac{3^d d}{\check{r}^d} \displaystyle\int\limits_{r-\check{r}}^{+\infty} \rho^{d-1}f(\rho) \,\mathrm{d}\rho }
\end{equation*}
\label{th:delSumInt}
\end{theorem}

\begin{proof}
Note that the condition $r \mapsto r^{d-1}f(r)$ is decreasing implies that $f$ itself is a decreasing function.
For a Delone set in $\mathbb{R}^d$ with covering radius $\check{r}$, the number of sites in the annulus with outer radius $(n+1)\check{r}$ and inner radius $n\check{r}$ is upper bounded by $\left(n+2\right)^d-\left(n-1\right)^d$ (see Annex \ref{an:numB}). Since $f$ is decreasing, it is upper bounded by $f(n\check{r})$ in this annulus. Given the positivity of $f$, the sum of the terms in this annulus is then upper bounded by:

\[
\displaystyle\sum_{\begin{subarray}{c} \rho \in \mathcal{L} \\ \floor{\frac{\rho}{\check{r}}} = n \end{subarray}}  f(\rho) \le \left(\left(n+2\right)^d-\left(n-1\right)^d\right)f(n\check{r})
\]

Considering that we want to sum over all the terms where $r \le \rho$, we obtain the following estimate:

\[
\sll f(\rho) \le \sum_{n\ge \floor{\frac{r}{\check{r}}}}f(n\check{r})\left(\left(n+2\right)^d-\left(n-1\right)^d\right)
\]

Since for $n\ge1$ we have $n+2\le 3n$ the term $\left(\left(n+2\right)^d-\left(n-1\right)^d\right)$ have $3^d d n^{d-1}$ as upper bound. The estimate for the sum thus becomes:

\[
\sll f(\rho) \le \frac{3^d d}{\hat{r}^{d-1}}  \sum_{n\ge \floor{\frac{r}{\check{r}}}} (n\hat{r})^{d-1} f(n\check{r})
\]

As the function $r \mapsto r^{d-1}f(r)$ is decreasing, we can obtain the upper estimate of the sum through the integral. Then considering the positivity of $r \mapsto r^{d-1}f(r)$. Expanding the domain of integration provides an additional upper estimate, given that $r-\check{r} < \floor{\frac{r}{\check{r}}}\check{r}$ we have:

\begin{equation*}
\boxed{
\sll f(\rho) \le \left(\frac{3}{\check{r}}\right)^d d \int_{r-\check{r}}^{+\infty} \rho^{d-1}f(\rho) \,\mathrm{d}\rho }
\end{equation*}

Addressing now the lower bound, consider the packing radius $\hat{r}$. In the annulus with outer radius $n\hat{r}$ and inner radius $(n-3)\check{r}$ the number of sites should be superior to $\left(\left(n-1\right)^d-\left(n-2\right)^d\right)$  (see Annex \ref{an:numB}). Since the function is decreasing it is lower bounded by $f(n\hat{r})$. Starting the sum at respectively $n=$3, 4, and 5 we find a lower bound of 3 times the sum on the Delone set:

\[
\frac{1}{3}\displaystyle\sum_{n\ge \floor{\frac{r}{\check{r}}}}\left(\left(n-1\right)^d-\left(n-2\right)^d\right)f(n\hat{r}) \le \sll f(\rho)
\]

\noindent Also, for $n\ge 3$ we have $ d\frac{n^{d-1}}{3^{d-1}} \le \left(n-1\right)^d-\left(n-2\right)^d$:

\[
\frac{\hat{r}d}{(3\hat{r})^d}\sum_{n\ge \floor{\frac{r}{\check{r}}}} (n\hat{r})^{d-1} f(n\hat{r}) \le \sll f(\rho)
\]

Due to the fact that $r \mapsto r^{d-1}f(r)$ is decreasing, we can similarly have a lower estimate for the sum by the integral and for $r \ge 3\hat{r}$. Also since $\floor{\frac{r}{\check{r}}} \hat{r} < r +\hat{r}$ and the function is positive we can remove part of the domain of integration to have another, smoother lower bound :

\begin{equation*}
\boxed{\frac{d}{(3\hat{r})^d}
\int_{r+\hat{r}}^{+\infty} \rho^{d-1}f(\rho) \,\mathrm{d}\rho \le \sll f(\rho)}
\end{equation*}

\end{proof}

\begin{remark}
Note that this implies that $\sum_{\rho \in \mathcal{L}} f(\rho)$ converges if and only if $\int_{0}^{+\infty} \rho^{d-1}f(\rho) \,\mathrm{d}\rho$ converges.

In the particular case where $f(r) = \frac{1}{r^\alpha}$, the sum $\sum_{\rho \in \mathcal{L}} \frac{1}{\rho^\alpha}$ converges if and only if $\alpha> d$  .
\end{remark}

%% file: globalBound.tex
We now utilize the definitions and results obtained thus far and analyze the case of power-law couplings, which is the main part of the paper. We show rigorously the conditions for obtaining Gaussian decay of the Ramsey profile, as a function of the power-law and the dimensionality. Specifically, we highlight the effect of nearby (strongly coupled) spins on the decay profile and its importance at low dimensions and for shorter-range interactions.

\begin{figure*}[tbh]
    \centering
    \includegraphics[width=\textwidth]{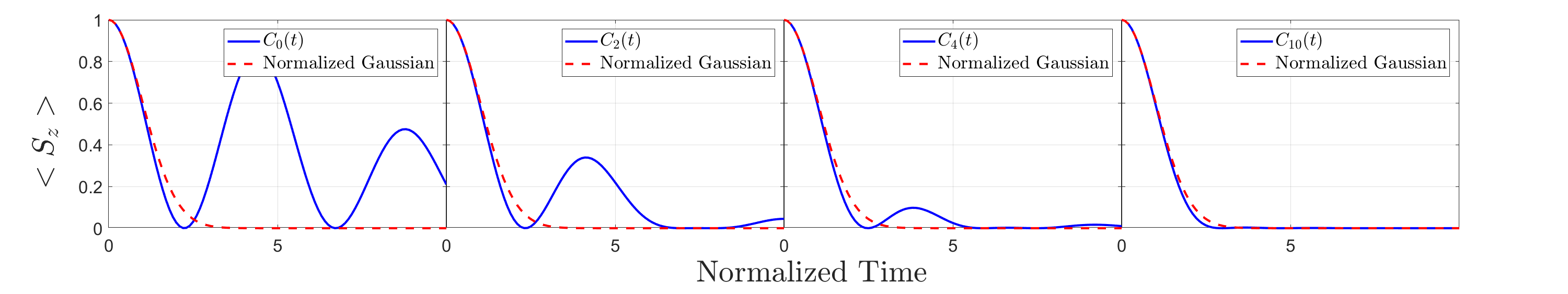}
    \caption{Convergence to a Gaussian by removing the close-by spins. Here the system is a 1D integer lattice and the coupling is of the form $A_i = \nicefrac{1}{r_i}$. The blue curve is the function $C_r$ previously defined and the red dashed curve is a fitted Gaussian. The horizontal axis is the rescaled time $\frac{t}{\sqrt{\sum A_i^2}}$. The vertical axis is the measured polarization of the central spin where 1 corresponds to a fully polarized state.}
\end{figure*}

In order to reduce the length of the calculation, we will utilize an asymptotic comparison notation.
For every function of two variables, we will note as shorthand $f(r,t)  \succeq  g(r,t)$ whenever $\exists k>0, \exists T>0,\exists R>0,(t>T \wedge r>R) \implies |f(r,t)| \geq k |g(r,t)|$. If we have $f\succeq g$ and $g \succeq f$ we will note $f \asymp g$.

Then note that in case of $f(\rho)=\frac{1}{\rho^\alpha}$ where $\alpha>d$ we can derive from theorem \ref{th:delSumInt} the following result:

\begin{equation}
\sll \frac{1}{\rho^\alpha} \asymp \frac{1}{r^{\alpha-d}}
\label{eq:mainResult}
\end{equation}

\subsection{Compact convergence}

\begin{theorem}
Let's have $\mathcal{L}$ a Delone set of $\mathbb{R}^d$, a function $A(r)=\frac{1}{r^\alpha}$ with $2\alpha > d$ and the following family of functions:  

\[
C_r(t) = \prod_{\begin{subarray}{l} \rho \in \mathcal{L}\\ \rho \ge r \end{subarray}} \cos\left(\frac{A(\rho) t}{\sqrt{\sum_{\begin{subarray}{l} \rho \in \mathcal{L}\\ \rho \ge r \end{subarray}} A(\rho)^2}}\right)
\]

\noindent Then this family converges compactly to $t\mapsto\me^{-\frac{t^2}{2}}$ as $r$ goes to the infinity.    
\label{th:compConv}
\end{theorem}

\begin{proof}
Using the Taylor theorem, for all real $t$ we have the following estimate $\me^{-\frac{t^2}{2}}-\cos(t) \le \frac{t^4}{12}$. The product is then bounded by:

\[
\left| \me^{-\frac{t^2}{2}}- C_r(t)\right|  \le \frac{t^4}{12}\frac{\sum_{\begin{subarray}{l} \rho \in \mathcal{L}\\ \rho \ge r \end{subarray}} \frac{1}{\rho^{4\alpha}}}{\left(\sum_{\begin{subarray}{l} \rho \in \mathcal{L}\\ \rho \ge r \end{subarray}}  \frac{1}{\rho ^{2\alpha}}\right)^2}. 
\]

Since $2\alpha > d$, we are ensured that the denominator and numerator converge. Using the estimate (\ref{eq:mainResult}) we deduce that:

\[
\frac{\sum_{\begin{subarray}{l} \rho \in \mathcal{L}\\ \rho \ge r \end{subarray}} \frac{1}{\rho^{4\alpha}}}{\left(\sum_{\begin{subarray}{l} \rho \in \mathcal{L}\\ \rho \ge r \end{subarray}}  \frac{1}{\rho ^{2\alpha}}\right)^2} \asymp \frac{1}{r^d}.
\]

Here the approximation does not depend on $t$, so there are two strictly positive constants $K$ and $R$ such that for all $r>R$:

\begin{equation}
\left|\me^{-\frac{t^2}{2}}- C_r(t)\right| \le K\frac{t^4}{r^d}.
\label{eq:compact}
\end{equation}

Take an arbitrary $T\in \mathbb{R}$, for $|t| \le T$ the limit where $r$ goes to the infinity is zero. This shows the compact convergence.  
\end{proof}

Nevertheless, compact convergence is not sufficient for our purpose, since it only provides an approximation of the dynamics on short timescales, and we are interested in the long-time dephasing properties of the system.

\subsection{Uniform decay at infinity}

We now turn to the analysis of the Ramsey decay at long times. First, we show that the rate of decay at infinity of the family of functions $C_r$ is independent of $r$. 

\begin{theorem}
Let's have $\mathcal{L}$ a Delone set of $\mathbb{R}^d$, a function $A(r)=\frac{1}{r^\alpha}$ with $2\alpha > d$ and the following family of functions:  

\[
C_r(t) = \prod_{\begin{subarray}{l} \rho \in \mathcal{L}\\ \rho \ge r \end{subarray}} \cos\left(\frac{A(\rho) t}{\sqrt{\sum_{\begin{subarray}{l} \rho \in \mathcal{L}\\ \rho \ge r \end{subarray}} A(\rho)^2}}\right)
\]

\noindent Then we have the following:

\begin{equation}
C_r(t) \preceq   \me^{-t^{\frac{d}{\alpha}}} 
\label{eq:uniform}
\end{equation}
\label{th:unifDecay}
\end{theorem}

\begin{proof}

Define $D_r(t) = -\log|C_r(t)|$ where the logarithm is continuously extended to the extended real line with $-\log(0) = \infty$.

To bound the cosines for all times recall that if $x<1$ then $|\cos(x)| \le \me^{-\frac{x^2}{2}}$ and otherwise $|\cos(x)| \le 1$.
Thus $-\log|\cos(x)|\ge \frac{x^2}{2}$ if $x<1$ and  $-\log|\cos(x)|\ge 0$ otherwise.

\noindent Then set $r(t) = \max(t^{\frac{1}{\alpha}},r)$ and $t_r = \frac{t}{\sqrt{\sum_{\rho>r}\frac{1}{\rho^{2\alpha}}}}$ we have:

\[
D_r(t) \ge \frac{t_r^2}{2}\sum_{\begin{subarray}{l} \rho \in \mathcal{L}\\ \rho \ge r(t_r) \end{subarray}} \frac{1}{\rho^{2\alpha}}.
\]

\noindent  Using the estimate (\ref{eq:mainResult}) we have:

\[
D_r(t) \succeq  \frac{t_r^2}{r(t_r)^{2\alpha-d}}.
\]

\noindent Plugging the expression for $r(t)$ we find:

\[
D_r(t) \succeq  \min\left(t_r^{\frac{d}{\alpha}},\frac{t_r^2}{r^{2\alpha-d}}\right).
\]

\noindent Now using again (\ref{eq:mainResult}) to estimate $t_r$:

\[
t_r \asymp  r^{\alpha-\frac{d}{2}} t
\]

\noindent Plugging this result into the previous equation we obtain two constant $A$ and $B$ such that:

\[
D_r\left(t
\right)
\ge 
\min\left( A t^{\frac{d}{\alpha}},
B t^2
\right).
\]

\noindent As the variable $t$ grows, the first argument will always become the minimum so the following result is obtained:

\[
D_r\left(t
\right)
\succeq
t^{\frac{d}{\alpha}}.
\]

\noindent Finally, going back to the expression for $C_r$ by exponentiation we obtain the desired result:

\[
C_r(t) \preceq   \me^{-t^{\frac{d}{\alpha}}}.
\]

\end{proof}

\noindent Note that related behavior has been identified in relevant experimental systems, e.g. in \cite{Davis2023} and references therein.

\subsection{Uniform bound}
Based on the previous result, we can now derive a bound uniform in time for the family of functions $C_r$ and prove its convergence to a Gaussian under the appropriate assumptions.

\begin{theorem}
Let's have $\mathcal{L}$ a Delone set of $\mathbb{R}^d$, a function $A(r)=\frac{1}{r^\alpha}$ with $2\alpha > d$ and the following family of functions:  

\[
C_r(t) = \prod_{\begin{subarray}{l} \rho \in \mathcal{L}\\ \rho \ge r \end{subarray}} \cos\left(\frac{A(\rho) t}{\sqrt{\sum_{\begin{subarray}{l} \rho \in \mathcal{L}\\ \rho \ge r \end{subarray}} A(\rho)^2}}\right)
\]

\noindent Then this family converge uniformly to $t\mapsto\me^{-\frac{t^2}{2}}$ as $r$ goes to infinity.

\noindent That is  $\lim_{r\rightarrow+\infty} \sup_{t\in\mathbb{R}}\left| 
C_r(t) - \me^{-\frac{t^2}{2}}
\right| = 0$.   
\end{theorem}

\begin{proof}
We begin with the inequality:

\begin{equation}
|C_r(t)-\me^{-\frac{t^2}{2}}| \le \epsilon.
\label{ineq}
\end{equation}

For an arbitrary $\epsilon$ the theorem \ref{th:unifDecay} tells us that for $T_1$ and $R_1$ such that $|t|>T_1$ and $r>R_1$  we have (\ref{ineq}). But theorem \ref{th:compConv} tells us that there is an $R_2$ such that for $|t|<T_1$ and $r>R_2$ we have (\ref{ineq}).
Thus, taking $R_3 = \max(R_1,R_2)$ for all $|t|\in \mathbb{R}$ and $r>R_3$ we have (\ref{ineq}).

\noindent This proves uniform convergence of $C_r$ to $t\mapsto\me^{-\frac{t^2}{2}}$ as $r\rightarrow \infty$, that is $\lim_{r\rightarrow\infty} \sup_{t\in\mathbb{R}}\left| 
C_r(t) - \me^{-\frac{t^2}{2}}
\right| = 0$. 
    
\end{proof}

This is the main result of the paper, confirming rigorously the Ramsey dephasing behavior, such that close-by spins give an oscillating part, while spins at larger distances give a Gaussian decay.

%% file: counterExample.tex
\section{Ramsey Dephasing with exponentially decaying coupling}

\begin{figure}[tbh]
    \centering
    \includegraphics[width=0.5\textwidth]{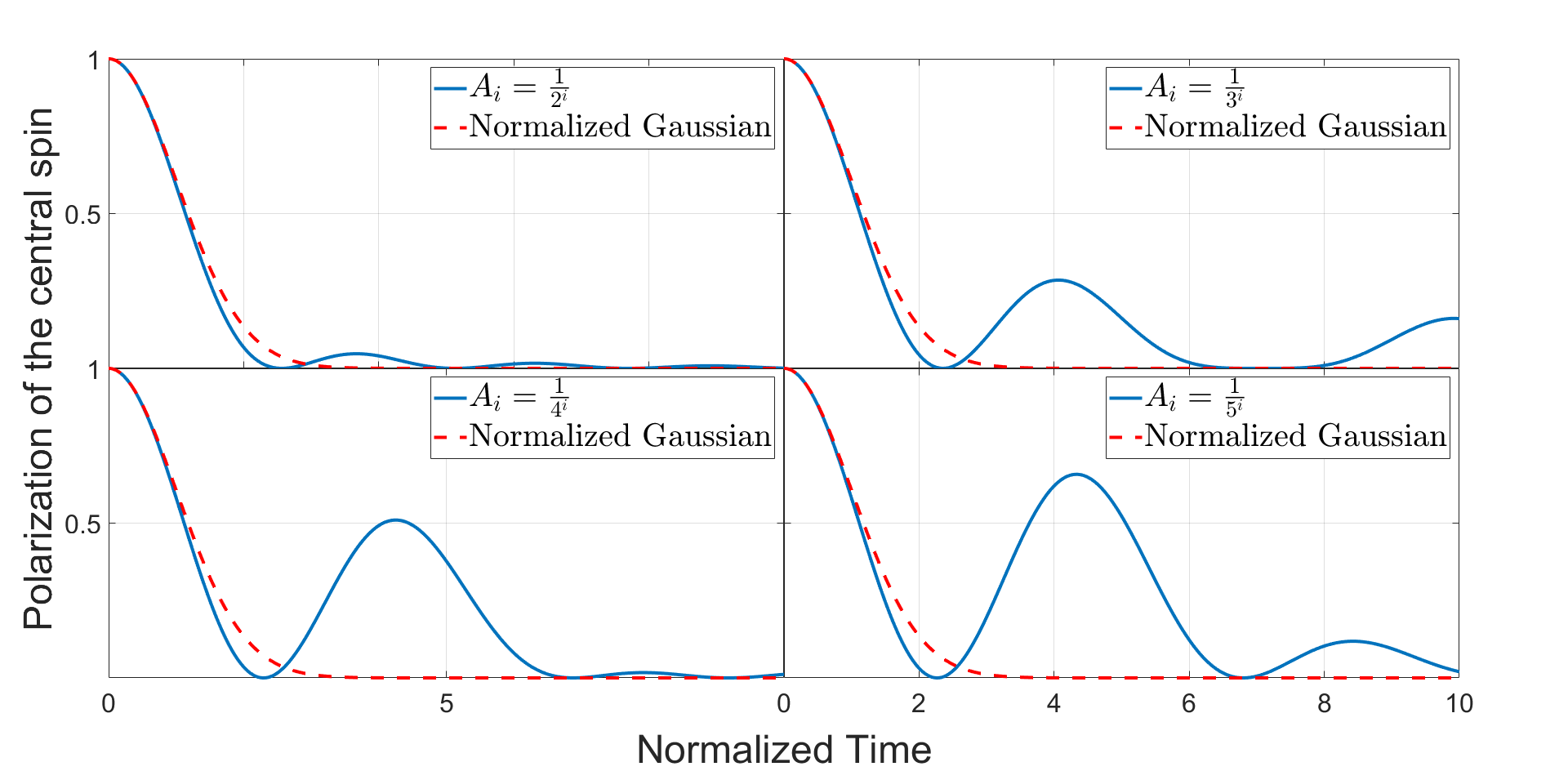}
    \caption{Illustration of non-Gaussian behavior in an exponentially decaying coupling system on a 1D integer lattice, where the coupling follows the form $A_i = \frac{1}{l^i}$ for some $l$. The blue curve represents the previously defined function $C$, while the red dashed curve shows a fitted Gaussian. The vertical axis denotes the measured polarization of the central spin, with 1 corresponding to a fully polarized state. The horizontal axis represents the rescaled time $\frac{t}{\sqrt{\sum A_i^2}}$. Removing nearby spins rescales the dynamics, ensuring the behavior remains unchanged on the rescaled time.}
    \label{fig:NG}
\end{figure}

We present counterexamples where the coupling exhibits exponential decay, demonstrating that in such cases, the behavior significantly deviates from Gaussian. This is evident from the fact that removing a single spin reduces the operator's size by a factor dependent on the decay length. As an illustration, Fig. \ref{fig:NG} shows different behaviors for couplings of the form $A_i = \frac{1}{l^i}$, where $l$ is an integer.

We first study an important case of exponentially decaying coupling of the form $A_i=\frac{1}{2^i}$ (where $i$ indices the bath spin position in the chain). 

Let's have a measure space $X$. For all $\phi$ in $L^{\infty}(X)$ we write $M_\phi$ as the multiplication operator by $\phi$. That is, for all $\psi$ in $L^2(X)$ we have $M_\phi \psi = \phi \psi$. The operator $M_\phi$ is then a bounded operator where $||M_\phi||=||\phi||_{\infty}$.

Note that we will talk of operators which are sum of $S_z^{(n)}$ terms so that we will talk only of the $x$ coordinate.

\begin{proposition} 

Let's have the following Hamiltonian :

\[
H = \displaystyle\sum_{k=1}^{+\infty}\frac{I^{(k)}_z}{2^k}
\]

\noindent Then $H$ is unitary equivalent to $M_x$ on $L^2([-1,1],\frac{\lambda}{2})$ where $\lambda$ is the Lebesgue measure.

\label{pr:sinc}
\end{proposition}

\begin{proof}
As seen in Annex \ref{an:Basis}, the ensemble of functions $\theta_\alpha$ for each $\alpha$ (a finite subset of $\mathbb{N}$) is an orthonormal basis of $L^2([-1,1],\frac{\lambda}{2})$. Similarly, we have defined for the same $\alpha$ the vector $\ket{\alpha}$ in $\mathcal{H}$ where for all $i\in \alpha$ the spin $i$ is pointing downward. By definition, it is an orthonormal basis of $\mathcal{H}$. The linear operator $U$ from $\mathcal{H}$ to $L^2([-1,1],\frac{\lambda}{2})$ defined by its action on the basis $U\ket{\alpha} = \theta_\alpha$ is then a unitary operator from $\mathcal{H}$ to $L^2([-1,1],\frac{\lambda}{2})$. 

 For all positive integers $n$ we have $U I^{(n)}_z U^{-1} = M_{\theta_{\{n\}}}$. Then $UHU^{-1}=M_f$, where $f=\sum_{k=1}^{+\infty}\frac{\theta_{\{k\}}}{2^k}$ the convergence of the series is understood in the $L^{\infty}([-1;1])$ topology.

Let's compute the Fourier coefficient $(\theta_{\{k\}})_n= \frac{1}{2}\int_{-1}^{1} \theta_{\{k\}}(x)\me^{-i\pi nx} \,\mathrm{d}x$ of the $\theta_{\{k\}}$ function.

First, after integration, the Fourier coefficients of $\theta_{\{1\}}$ are $\left(\theta_{\{1\}}\right)_{2n+1}=-\frac{i}{\pi}\frac{1}{n+\frac{1}{2}}$ and $\left(\theta_{\{1\}}\right)_{2n}= 0$. The action of $T$ (defined in Annex \ref{an:Basis}) in the Fourier basis is $(Tf)_{2n}= (-1)^n f_n$ and $(Tf)_{2n+1}= 0$. By applying multiple times the $T$ operator we have for all integers $n$ and positive integers $k$ greater than $1$ the equality $(\theta_{\{k\}})_{2^{k}n+2^{k-1}}=\frac{i}{\pi}\frac{1}{n+\frac{1}{2}}$, with all other coefficients being zero.

Eventually, for $k>1$ we have $\left( \frac{\theta_{\{k\}}}{2^k} \right)_{2^{k}n+2^{k-1}}=\frac{i}{\pi}\frac{1}{2^{k}n+2^{k-1}}$ and for $k=1$ we have $\left( \frac{\theta_{\{1\}}}{2} \right)_{2n+1}=-\frac{i}{\pi}\frac{1}{2n+1}$.

Let's have $a\in \mathbb{Z}\setminus \left\{0\right\}$. By the fundamental theorem of arithmetic, there is a unique positive integer $k$ such that $a=2^{k-1} b$ where $b$ is odd. Then there is a unique integer $n$ such that $b=2n+1$. We conclude that we can write in a unique way each $a\in \mathbb{Z}\setminus \left\{0\right\}$ in a form $a=2^{k-1}(2n+1)=2^{k}n+2^{k-1}$ with $k\in \mathbb{N}$ and $n\in \mathbb{Z}$.
 
The Fourier coefficients $\left(\sum_{k=1}^{+\infty} \frac{\theta_{\{k\}}}{2^k}\right)_n$ are thereby $(-1)^n\frac{ i}{\pi n}$ for $n\neq 0$ and $0$ for $n=0$. These are the Fourier coefficients of the $x$ function and having the same Fourier coefficients implies that $\sum_{k=1}^{+\infty} \frac{\theta_{\{k\}}}{2^k}$ is equal almost everywhere to $x$, implying furthermore that $U H U^{-1} = M_x$. 
\end{proof}

Let's have a central spin in contact with an infinite bath of spins with the following Hamiltonian:

\[
H = \frac{1}{2}\displaystyle\sum_{k=1}^{+\infty}\frac{S_z I^{(k)}_z}{2^k}
\]

\noindent We have $H=x\frac{S_z}{2}$ and $\me^{-ixt\frac{S_z}{2}}=\cos(\frac{xt}{2})-i\sin(\frac{xt}{2})S_z$.

\noindent For a general state $\rho=\frac{1+\Vec{v}(0)\cdot\Vec{S}}{2}$ of the central spin we have $U\left(\frac{1+\Vec{v}(0)\cdot\Vec{S}}{2}\right) U^\dagger=\frac{1+\Vec{v}(t)\cdot\Vec{S}}{2}$ where:

\[
\begin{cases}
v_x(t)=& \cos(xt)v_x(0)-\sin(xt)v_y(0) \\
v_y(t)=& \sin(xt)v_x(0) + \cos(xt) v_y(0) \\
v_z(t)=& v_z(0)
\end{cases}
\]

Now assume the bath is in a fully polarized state, that is the indicator function is $[-1;1]$. Then for tracing out the bath we get $\bra{\emptyset}\cos(xt))\ket{\emptyset}=\int_{-1}^1\cos(xt)\,\frac{\mathrm{d}x}{2} = \mathrm{sinc}(t)$ and  $\bra{\emptyset}\sin(xt))\ket{\emptyset}=\int_{-1}^1\sin(xt)\,\frac{\mathrm{d}x}{2}=0$. Thus we obtain:

\[
\begin{cases}
v_x(t)=& \mathrm{sinc}(t)v_x(0)\\
v_y(t)=& \mathrm{sinc}(t)v_y(0) \\
v_z(t)=& v_z(0)
\end{cases}
\]

This is pure dephasing of the spin and removing close-by spins will not make the operator closer to a Gaussian.

\noindent From proposition \ref{pr:sinc} the following can be deduced for a slightly different exponentially decaying coupling form:

\begin{proposition} 

Let’s have the following Hamiltonian :

\[
H = \displaystyle\sum_{k=1}^{+\infty}\frac{I^{(k)}_z}{3^k}
\]

\noindent Then $H$ is unitary equivalent to the multiplication operator by $x-\frac{1}{2}$ on the space $L^2([0,1],\mu)$ where $\mu$ is the Cantor distribution.
\label{pr:Cantor}
\end{proposition}

Indeed, using again the set of functions from Annex \ref{an:Basis}, the operator $H$ is unitary equivalent to the multiplication operator by $\sum_{n=1}^\infty \frac{\theta_{\{n\}}(x)}{3^n}$ on the space $L^2([-1,1],\frac{\lambda}{2})$.

The Cantor set $\mathcal{C}$ is defined to be the set of numbers of the form $\sum_{n=1}^\infty \frac{2a_n}{3^n}$, where each $a_n$ is either $0$ or $1$. Let's denote $C$ as the Cantor ternary function. On the Cantor set the Cantor function acts by definition in the following way:

\[
C\left(\sum_{n=1}^\infty \frac{2a_n}{3^n}\right)=\sum_{n=1}^\infty \frac{a_n}{2^n}
\]

The set $\mathcal{E}$ of endpoints of the Cantor set is defined to be the points $\sum_{n=1}^\infty \frac{2a_n}{3^n}\in\mathcal{C}$, where there is an $N$ such that either $a_n=0$ for all $n$ that are bigger than $N$ or $a_n=1$ for all $n$ that are bigger than $N$.
The Cantor function is a bijection from $\mathcal{C}\setminus\mathcal{E}$ to $[0,1]\setminus\mathcal{D}$, and we will denote $C^{-1}$ its inverse on this domain. Then the following proposition is derived:

\begin{proposition} For all $x\in[-1,1]\setminus\mathcal{D}$ we have:

\[
\sum_{n=1}^\infty \frac{\theta_{\{n\}}(x)}{3^n}=C^{-1}\left(\frac{x+1}{2}\right)-\frac{1}{2}
\]
\end{proposition}

\begin{proof} For $x\in [-1,1]\setminus\mathcal{D}$ we can remark that $\displaystyle\sum_{n\in \mathbb{N}} \frac{\theta_{\{n\}}(x)}{3^n}+\frac{1}{2}=\sum_{n\in \mathbb{N}}\frac{2\left(\frac{\theta_{\{n\}}(x)+1}{2}\right)}{3^n}$ and $\frac{\theta_{\{n\}}(x)+1}{2}\in\{0,1\}$.

\noindent So applying the Cantor function we get $C\left(\sum_{n=1}^\infty \frac{\theta_{\{n\}}(x)}{3^n}+\frac{1}{2}\right)=\sum_{n\in \mathbb{N}}\frac{\frac{\theta_{\{n\}}(x)+1}{2}}{2^n}=\frac{x+1}{2}$.

\noindent However, since the Cantor function is bijective on $\mathcal{C}\setminus\mathcal{E}$ we have $\sum_{n=1}^\infty \frac{\theta_{\{n\}}(x)}{3^n}+\frac{1}{2}=C^{-1}\left(\frac{x+1}{2}\right)$.
\end{proof}

The Cantor function is not bijective but we can obtain the following result:

\begin{proposition} 

Let's $D(x)=\sum_{n=1}^\infty \frac{\theta_{\{n\}}(2x-1)}{3^n}+\frac{1}{2}$ and $\mu$ the image measure of $\lambda$ by $D$ we have:

\begin{enumerate}
   \item $\mu$ is the Cantor distribution
   \item $C(D(x))=x\text{ $\lambda$-almost everywhere}$
   \item $D(C(x))=x\text{ $\mu$-almost everywhere}$
\end{enumerate}

\end{proposition}

\begin{proof}
Since $\lambda$ is finite $\mu$ is finite and we can deduce it from its Fourier transform.

We have $\bra{\emptyset}\me^{it(B+\frac{1}{2})}\ket{\emptyset}=\int_\mathbb{R} \me^{itx} \,\mathrm{d}\mu(x)$.

But $\bra{\emptyset}\me^{it(H+\frac{1}{2})}\ket{\emptyset}=\me^{\frac{it}{2}}\prod_{n=1}^\infty\cos\left(\frac{t}{3^n}\right)$ and this function is known to be the characteristic function of the Cantor distribution. 

We know that $\mathcal{D}$ is a negligible space for $\lambda$ and that for $x\in[0,1]\setminus\mathcal{D}$ we have $D(x)=C^{-1}(x)$, so that $C(D(x))=x$ $\lambda$-almost everywhere.

Similarly we have $\mu(\mathcal{C}\setminus\mathcal{E})=\lambda({D}^{-1}(\mathcal{C}\setminus\mathcal{E}))=\lambda([0,1]\setminus\mathcal{D})=1$ but $\mu(\mathcal{C})=1$, so that $\mu(\mathcal{E})=0$ and we have $D(C(x))=x$ $\mu$-almost everywhere.   
\end{proof}

\begin{proposition} 

Let's $(U\phi)=\phi(2C(x)-1)$, then $U$ is an unitary operator from $L^2([-1,1],\frac{\lambda}{2})$ to $L^2([0,1],\mu)$.

\end{proposition}
\begin{proof}

We have $\braket{U\phi}{U\psi}=\int_0^1 \overline{\phi(2C(x)-1)}\psi(2C(x)-1)\,\mathrm{d}\mu(x)$, but since $\mu$ is the Cantor distribution 
$\braket{U\phi}{U\psi}=\int_0^1 \overline{\phi(2x-1)}\psi(2x-1)\,\mathrm{d}x=
\int_{-1}^1 \overline{\phi(x)}\psi(x)\,\frac{\mathrm{d}\lambda}{2}=\braket{\phi}{\psi}$.
So $U$ is an isometry from $L^2([-1,1],\frac{\lambda}{2})$ to $L^2([0,1],\mu)$

Now let's have $f$ measurable, and since $D$ is a measurable function $f\circ D$ is measurable. Then we have $f\circ D\circ C = f$ $\mu$-almost everywhere.

So we have shown that if $f$ is measurable there is a $g$ which is measurable such that $f=g\circ C$ $\mu$-almost everywhere.

The previous result tells us that if $g$ is measurable there is an $f$ which is measurable such that $g=Uf$ $\mu$-almost everywhere so that $U$ is subjective. Furthermore, since $U$ is an isometry $U$ is a unitary operator.

\end{proof}

\begin{proof}[Proof of proposition \ref{pr:Cantor}]

For all $\psi$ and $\phi$ in $L^2([-1,1],\frac{\lambda}{2})$ we have $\bra{\phi}H+\frac{1}{2}\ket{\psi}=\int_0^1 D(x) \overline{\phi(2x-1)}\psi(2x-1)\,\mathrm{d}x$.

But since $C\left(D(x)\right)=x$ almost everywhere we have:
\begin{multline*}
\bra{\phi}H+\frac{1}{2}\ket{\psi}= \\ \int_0^1 D(x)\overline{\phi\left(2C\left(D(x)\right)-1\right)}\psi\left(2C\left(D(x)\right)-1\right)\,\mathrm{d}x
\end{multline*}

And by definition of the image measure, we have:

\[
\bra{\phi}B+\frac{1}{2}\ket{\psi}
=\int_0^1 x \overline{\phi(2C(x)-1)}\psi(2C(x)-1)\,\mathrm{d}\mu(x)
\]

\noindent But by definition of $U$:

\[
\bra{\phi}H+\frac{1}{2}\ket{\psi}
=\bra{\phi}U^\dagger M_x U\ket{\psi}
\]

Since $\phi$ and $\psi$ are arbitrary vectors we deduce that $H= U^\dagger  M_{x-\frac{1}{2}} U$.
\end{proof}

Now the Cantor distribution is known to be an example of a singular continuous distribution.
That is, it doesn't possess a probability density function or probability mass function. Since $B$ is unitary equivalent to the multiplication operator by $x-\frac{1}{2}$ on the space $L^2([0,1],\mu)$, its spectrum is the support of $\mu$ that is $\sigma(B) = \mathcal{C}-\frac{1}{2}$. This shows that for this coupling form the dephasing does not converge to a Gaussian, yet displays a fractal, singular continuous form. 

One of the consequences of a singular continuous spectrum is that it allows the dynamic to not converge to $0$ at infinity. We observe that this is indeed the case. The dynamics of the polarization of the central spin are described by the quantity $C(t) = \prod_{k=1}^\infty \cos\left(\frac{t}{3^k}\right)$. 
Since the series $\sum_{k=1}^\infty \sin^2\left(\frac{\pi}{3^k}\right)$ converges, we have $C(\pi) = \prod_{k=1}^\infty \cos\left(\frac{\pi}{3^k}\right) = l$, where $l > 0$. Numerical calculations give $l \approx 0.46$, indicating a nearly half-polarized spin. Letting $ t_i = 3^i \pi $, we have $C(t_i) = (-1)^i l$, and since $t_i \rightarrow +\infty$, this demonstrates that $ C(t) $ does not converge to a limit as $ t $ approaches infinity. Interestingly, using the same reasoning we can show that for every integer $n>2$ the function $\prod_{k=1}^\infty \cos(\frac{t}{n^k})$ does not go to zero at infinity. A general result, as demonstrated in \cite{salem}, shows that if $l\neq 2$ then $ C(t) = \prod_{k=1}^\infty \cos\left(\frac{t}{l^k}\right) $ approaches zero as $ t $ tends to infinity if and only if $ l $ is not a Pisot number. In \cite{salem}, Pisot numbers are referred to as numbers of class S.

This section illustrates how dephasing can differ from Gaussian decay, even when neighboring spins are removed. The coupling $A_i = \frac{1}{2^i}$ shows a decaying behavior following a $\frac{1}{t}$ pattern. More notably, the example with the coupling $A_i = \frac{1}{3^i}$ demonstrates a spectrum that is not absolutely continuous but rather singular continuous, exhibiting fractal behavior and resulting in the absence of total dephasing and the presence of persistent oscillations.

\begin{figure}[tbh]
    \centering
    \includegraphics[width=0.5\textwidth]{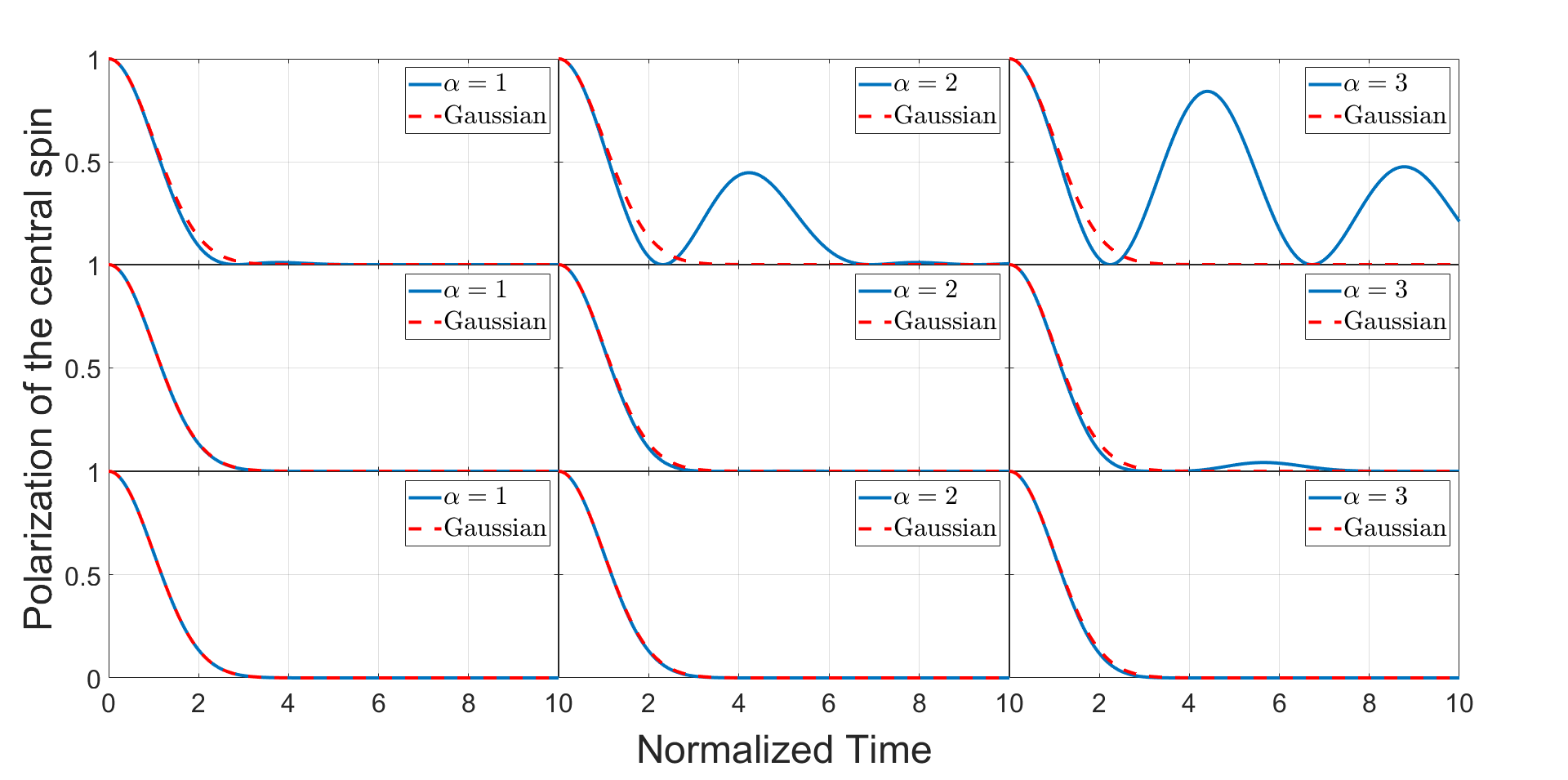}
    \caption{The behavior of the central spin is plotted for different exponents $\alpha$ and dimensions on an integer lattice, where the coupling is given by $A_i = \frac{1}{r_i^\alpha}$. The blue curve corresponds to the previously defined function $C$, while the red dashed curve represents a fitted Gaussian. The vertical axis shows the measured polarization of the central spin, with 1 indicating a fully polarized state. The horizontal axis represents the rescaled time $\frac{t}{\sqrt{\sum A_i^2}}$. For $\alpha = 1$, the behavior of the central spin is approximately Gaussian, even with nearby spins. However, in 1D for $\alpha = 2, 3$, oscillations dominate, and many nearby spins must be removed to approach a Gaussian distribution.}
    \label{fig:cases}
\end{figure}

%% file: conclusion.tex
In this article, we have considered a system comprising a central spin-$\frac{1}{2}$ interacting with an unpolarized spin-$\frac{1}{2}$ bath through an inverse power-law Ising coupling $A_i = \frac{1}{r_i^\alpha}$, with $r_i$ being the distance between the central spin and bath spin $i$. We have formally established that removing nearby spins (a limited, finite set) causes the dephasing of the central spin to converge to a Gaussian decay across various configurations (namely, the Delone set---i.e., homogeneous disorder) and for all exponents $\alpha > \frac{d}{2}$, with $d$ being the dimensionality. The final decay will be approximately the product of an oscillating function and a Gaussian, the approximation being uniform in time. This study confirms the initial assumption of Gaussian behavior for an inverse power-law decaying interaction but also reveals the limitations imposed by the presence of nearby spins.

The results of this study are primarily qualitative, leaving open the question of how many nearby spins must be removed to achieve a desired precision. As illustrated in Figure \ref{fig:cases}, in ordered systems, oscillations become more pronounced at higher values of the exponent and in smaller dimensions. A deeper understanding of how the number of spins to remove scales with dimension, exponent, and other system characteristics could provide valuable insights into generating dephasing patterns that deviate from Gaussian behavior, potentially in experimentally relevant scenarios.

The calculations performed for the inverse power-law coupling are not applicable to scenarios involving exponential decay. We have shown explicitly that, in the case of couplings $\frac{1}{2^i}$, the dynamics are described by a cardinal sine, and that for couplings $\frac{1}{3^i}$, the spectrum is singularly continuous. One consequence of the latter is the presence of persistent oscillations, even in systems with an infinite number of spins. Investigating the implications of a singular continuous spectrum in the Hamiltonian remains an intriguing direction for future research.

%% file: anexe/anexe.tex
\part*{Annexes}
\setcounter{section}{0}

\subfile{"./numberInAnnulus"}

\subfile{"./integralSumEstimate"}

\subfile{"./basisFunction"}

\subfile{"./representation"}

%% file: anexe/numberInAnnulus.tex
\section{Annulus number of site estimate} 
\label{an:numB}

\subsection{Upper-bound}

Take a annulus with outer radius $b$ and inner radius $a$ where we consider $a\ge \check{r}$ centered at zero and noted $\mathrm{Ann}(a,b)$.
Then the union of the balls with radius $\check{r}$ and centered at the points of $\mathcal{L}\cap\mathrm{Ann}(0;a,b)$ is covered by $\mathrm{Ann}(a-\check{r},b+\check{r})$.

Furthermore since the balls does not intersect their volume is $N \lambda(B(r,\check{r}))$ so we deduce that $N \lambda(B(r,\check{r})) \le \lambda(B(b+\check{r}))-\lambda(B(a-\check{r}))$. Remember that $\frac{\lambda(B(r,a))}{\lambda(B(r,\check{b}))}=\left(\frac{a}{b}\right)^d$ we get:

\[
N \le 
\left(\frac{b}{\check{r}}+1\right)^d
-\left(\frac{a}{\check{r}}-1\right)^d
\]

\subsection{Lower-bound}

On the other side, the union of the closed balls with radius $\hat{r}$ and centered at the points of $\mathcal{L}\cap\overline{\mathrm{Ann}}(0;a,b)$ cover the closed annulus of outer radius $b-\hat{r}$ and inner radius $a+\hat{r}$.

Indeed if not, there would be $y\in \overline{\mathrm{Ann}}(b-\hat{r},a+\hat{r})$ such that $||y-x||> \hat{r}$ for $x\in\mathcal{L}\cap\overline{\mathrm{Ann}}(0;a,b)$.
But if $x\in \mathcal{L}\setminus\overline{\mathrm{Ann}}(0;a,b)$ then $||y-x||>\hat{r}$. Hence $||y-x||\ge \hat{r}$ for all $x\in\mathcal{L}$ however by definition each point of the space should have a point in $\mathcal{L}$ closer than $\hat{r}$, contradiction.

So that the area $\lambda(B(b-\hat{r}))-\lambda(B(a+\hat{r})) \le N\lambda(B(r,\hat{r}))$ and we have:

\[
\left(\frac{b}{\hat{r}}-1\right)^d
-\left(\frac{a}{\hat{r}}+1\right)^d
\le N 
\]

In the end we get the estimate:

\[
\left(\frac{b}{\hat{r}}-1\right)^d
-\left(\frac{a}{\hat{r}}+1\right)^d
\le N \le 
\left(\frac{b}{\check{r}}+1\right)^d
-\left(\frac{a}{\check{r}}-1\right)^d
\]

%% file: anexe/integralSumEstimate.tex
\section{Sum-Integral estimate}
\label{an:smInt}

Suppose we have $f$ a decreasing function then for all $x\in[n,n+1]$ we have $f(x) \le f(n) \le f(x-1)$ and after integration $\int_n^{n+1} f(x) \,\mathrm{d}x\le f(n) \le \int_{n-1}^{n} f(x) \,\mathrm{d}x$ so that after summation:

\[
\int_a^\infty f(x) \,\mathrm{d}x \le \sum_{n=a}^\infty f(n) \le \int_{a-1}^\infty f(x) \,\mathrm{d}x
\]

%% file: anexe/basisFunction.tex
\section{A Hilbert basis of function} 
\label{an:Basis}

Let $\mathcal{F}$ denote the collection of finite subsets of natural numbers. Here, we establish a specific Hilbert basis on $L^2\left([-1,1],\frac{\mathrm{d} x}{2}\right)$, which is parameterized by $\mathcal{F}$.

We define functions over $[-1,1]$: $\theta_\emptyset$ as the indicator function of $[-1,1]$, denoted by $\mathbf{1}_{[-1,1]}$, and $\theta_{\{1\}}$ as the sign function, symbolized by $\sigma$.

An operator $T$ is defined by his action on $\psi$ as $(T\psi)(x)=\psi\left(2x-\sigma(x)\right)$, and $\theta_{\{n+1\}}$ is defined as $T\theta_{\{n\}}$.

Eventually, for any finite subset $\alpha$ of $\mathbb{N}$, we define functions $\theta_{\alpha}$ as the product over $n$ in $\alpha$ of $\theta_{\{n\}}$. This establishes a set of functions in $L^2\left([-1,1],\frac{\mathrm{d} x}{2}\right)$ indexed by $\mathcal{F}$.

\begin{proposition*}
The collection of functions $(\theta_\alpha)_{\alpha\in\mathcal{F}}$ forms an orthonormal basis for $L^2\left([-1,1],\frac{\mathrm{d} x}{2}\right)$.
\end{proposition*}

\begin{proof}

We demonstrate that $(\theta_\alpha)_{\alpha\in\mathcal{F}}$ is first an orthonormal set and second a total set.

\proofpart{1}{Orthonormal set}

Because $\theta_{\{n\}}^2=1$ for all $n\in \mathbb{N}$, it follows that $\theta_{\alpha\Delta\beta}=\theta_{\alpha}\theta_{\beta}$ for all finite subsets $\alpha$ and $\beta$ of the natural numbers, where $\alpha\Delta\beta$ denotes the symmetric difference of $\alpha$ and $\beta$.

From this, we infer that $\braket{\theta_\alpha}{\theta_\beta}=\int_{-1}^1 \theta_{\alpha}(x)\theta_{\beta}(x) \,\frac{\mathrm{d}x}{2}=\int_{-1}^1 \theta_{\alpha\Delta\beta} \,\frac{\mathrm{d}x}{2}$. Thus, evaluating the integral $\int_{-1}^1 \theta_{\alpha} ,\frac{\mathrm{d}x}{2}$ for all $\alpha$ will enable us to verify whether it forms an orthonormal set.

Firstly, observe that $\int_{-1}^1 \theta_\emptyset(x) \,\frac{\mathrm{d}x}{2} = 1$ so we can restrict to $\alpha\neq\emptyset$. Furthermore, a change of variable shows that $\int_{-1}^1 (T\psi)(x) \,\mathrm{d}x = \int_{-1}^1 \psi(x) \,\mathrm{d}x$. Additionally, it's evident that $T(\psi\phi) = (T\psi)(T\phi)$. Denote the set $\alpha+n$ as $\{{m+n | m\in \alpha }\}$ it follows that $T\theta_\alpha=\theta_{\alpha+1}$. Based on the invariance of the integral under the action of $T$, for all non-empty $\alpha$, there exists a $\beta$ such that $1\in\beta$ and $\int_{-1}^1 \theta_{\alpha} \,\frac{\mathrm{d}x}{2}=\int_{-1}^1 \theta_{\beta} \,\frac{\mathrm{d}x}{2}$. Hence, we can suppose without loss of generality that $1\in\alpha$.

Given that $1\in\alpha$ we have by definition $\theta_\alpha=\sigma\theta_{\alpha\setminus\{1\}}$. Then $\int_{-1}^1 \sigma(x)\theta_{\alpha\setminus\{1\}}(x) \,\frac{\mathrm{d}x}{2} = \frac{1}{2}\int_{-1}^1 \left[\theta_{\alpha\setminus\{1\}}\left(\frac{x+1}{2}\right)-\theta_{\alpha\setminus\{1\}}\left(\frac{x-1}{2}\right)\right] \,\mathrm{d}x$. But, since by definition $1\notin \alpha\setminus\{1\}$, there exists a $\beta$ such that $T\theta_\beta=\theta_{\alpha\setminus\{1\}}$. However, $(T\theta_\beta)\left(\frac{x\pm 1}{2}\right) = \theta_\beta\left(x\pm 1-\sigma\left(\frac{x\pm 1}{2}\right)\right)$, and since for $x\in[-1,1]$, we have $\sigma\left(\frac{x\pm 1}{2}\right) = \pm 1$, it follows that $\theta_\beta\left(x \pm 1-\sigma\left(\frac{x\pm 1}{2}\right)\right)=\theta_\beta(x)$. Therefore, for all non-empty $\alpha$, we have $\int_{-1}^1 \theta_\alpha(x) \,\frac{\mathrm{d} x}{2} = 0$.

Eventually, we have shown that first $\int_{-1}^1 \theta_\emptyset(x) \,\frac{\mathrm{d}x}{2} = 1$ and second for all $\alpha \neq \emptyset$ we have $\int_{-1}^1 \theta_\alpha(x)\,\frac{\mathrm{d} x}{2}=0$. Since $\alpha\Delta\beta=\emptyset$ if and only if $\alpha=\beta$ we deduce that $\braket{\theta_\alpha}{\theta_\beta}=\delta_{\alpha\beta}$. This show that $(\theta_\alpha)_{\alpha\in \mathcal{F}}$ is an orthonormal set.    

\proofpart{2}{Total set}

We aim to demonstrate that $(\theta_\alpha)_{\alpha\in\mathcal{F}}$ constitutes a total set, meaning that the linear span of $(\theta_\alpha)_{\alpha\in\mathcal{F}}$ is dense in $L^2\left([-1,1],\frac{\mathrm{d} x}{2}\right)$.

Note the multiplication operator by a function $f$ as $M_f$ and observe that the linear span of $(\theta_\alpha)_{\alpha\in\mathcal{F}}$ remains unchanged under the action of $T$, $\frac{1+M_\sigma}{2}$ and $\frac{1-M_\sigma}{2}$.

Consider the functions $\frac{\theta_{\emptyset}+\theta_{{1}}}{2} = \mathbf{1}_{[0,1]}$ and $\frac{\theta_{\emptyset}+\theta_{{1}}}{2} = \mathbf{1}_{[-1,0]}$.

Then, for $k\in [2^{n+1},2^{n+1}-1]$, we have $\left(\frac{1+M_\sigma}{2}\right)T\mathbf{1}_{\left[\frac{k}{2^n},\frac{k+1}{2^n}\right]}=\mathbf{1}_{\left[\frac{k+2^n}{2^{n+1}},\frac{k+1+2^n}{2^{n+1}}\right]}$ and $\left(\frac{1-M_\sigma}{2}\right)T\mathbf{1}_{\left[\frac{k}{2^n},\frac{k+1}{2^n}\right]}=\mathbf{1}_{\left[\frac{k-2^n}{2^{n+1}},\frac{k+1-2^n}{2^{n+1}}\right]}$. In other words, for all $k\in [2^{n+1},2^{n+1}-1]$, the functions $\mathbf{1}_{\left[\frac{k}{2^{n+1}},\frac{k+1}{2^{n+1}}\right]}$ belong to the span of $(\theta_\alpha)_{\alpha\in\mathcal{F}}$. 

By induction and linear combination, for all natural numbers $n$ and for $a < b$ in $[2^{n+1},2^{n+1}-1]$, the function $\mathrm{1}_{\left[\frac{a}{2^n},\frac{b}{2^m}\right]}$ belongs to the span of $(\theta_\alpha)_{\alpha\in\mathcal{F}}$.

Consider the sequence of functions $f_n=\mathbf{1}_{\left[\frac{\floor{2^n x}}{2^n},\frac{\ceil{2^n y}}{2^n}\right]}$ for all $x$ and $y$ in $[-1,1]$ such that $-1\le x < y \le 1$. Then, for all natural integers $n$, the function $f_n$ is within the span of $(\theta_\alpha)_{\alpha\in\mathcal{F}}$. 

But, as $n$ goes to the infinity, the sequence $f_n$ converges to $\mathbf{1}_{[x,y]}$ in the $L^2\left([-1,1],\frac{\mathrm{d} x}{2}\right)$ topology, since $||f_n-\mathrm{1}_{[x,y]}||^2 = \frac{1}{2}\left(x-\frac{\floor{2^n x}}{2^n}+\frac{\ceil{2^n y}}{2^n}-y\right) \le \frac{1}{2^n}$.

Therefore, the indicator functions of intervals are in the closure of the span of $(\theta_\alpha)_{\alpha\in\mathcal{F}}$. Since the set of indicator functions of intervals forms a total set of $L^2([-1, 1], \frac{\mathrm{d} x}{2})$, it follows that $(\theta_\alpha)_{\alpha\in\mathcal{F}}$ is also a total set.

Having shown that $(\theta_\alpha)_{\alpha\in\mathcal{F}}$ is both an orthonormal and total set, it forms an orthonormal basis for $L^2\left([-1,1],\frac{\mathrm{d} x}{2}\right)$.

\end{proof}

%% file: anexe/representation.tex
\section{GNS construction of unpolarized state}
\label{an:GNS}

\subsection{Definition of the unpolarized state}

For all bounded open subset $O$ of $\mathbb{R}^d$ define the state $\tau$ on $\mathfrak{U}_O\subset\mathfrak{U}$ by $\tau: a\mapsto \frac{1}{\dim(\mathfrak{U}_O)}\Tr_{\mathfrak{U}_O}(a)$.

It define a continuous linear form on $\mathfrak{U}_\text{local}$ which, by Hahn–Banach theorem, has a unique extension to a continuous linear form on $\mathfrak{U}$ with same norm.

That's the unique state such that for every pair of elements of the Quasi-local algebra we have $\tau(ab)=\tau(ba)$. And it is interpreted as the unpolarized state of the spin bath.

\subsection{GNS construction}

The only element where $\tau(a^* a) = 0$ is the zero element. So the Hilbert space $\mathcal{H}_\tau$ associated to the state $\tau$ is the completion of the vector space associated to $\mathfrak{U}$ with respect to the norm $||a||_\tau = \sqrt{\tau(a^*a)}$ and scalar product $\braket{a}{b}=\tau(a^* b)$.

We have an linear isometry $\Phi$ from $\mathfrak{U}$ to a dense subspace of $\mathcal{H}_\tau$.
Furthermore, for all $a\in\mathfrak{U}$ we have $||\Phi(a)||_\tau \leq ||a||$ meaning that $\Phi$ is continuous from the topology of $\mathfrak{U}$ to the topology of $\mathcal{H}_\tau$

We have $\mathcal{H}_\tau$ defined as the closure of $\Phi(\mathfrak{U})$ but since $\mathfrak{U}_\mathrm{loc}$ is dense in $\mathfrak{U}$ and $\Phi$ is continuous we have $\Phi(\mathfrak{U}_\mathrm{loc})$ dense in $\mathcal{H}_\tau$. 

Note $\Gamma$ the ensemble of finite subset of $\mathcal{L}\times\{1,2,3\}$ and let $\alpha\in\Gamma$ we note $S_\alpha = \prod_{x\in \alpha} S_{\alpha_2}^{(\alpha_1)}$ were $S_i^{(x)}$ are the Pauli matrices at site $x$ in the lattice for $i\in\{1,2,3\}$. Then for all $\alpha\in\Gamma$ we have $S_\alpha$ is in $\mathfrak{U}_\text{local}$. Furthermore, its a basis of the local algebra and we see that $\tau(S_\alpha S_\beta) = \delta_{\alpha\beta}$. So that the family $(S_\alpha)_{\alpha\in \Gamma}$ is orthonormal and generate the local algebra which is dense in $\mathcal{H}_\tau$ so by definition it's an Hilbert basis of $\mathcal{H}_\tau$.

For all $v\in \mathcal{H}_\tau$ we have $v = \sum_{\alpha\in\Gamma} \tau(S_\alpha v)S_\alpha$. And $\mathcal{H}_\tau$ is unitary equivalent to $l^2(\Gamma)$ with the unitary $v \mapsto (\tau(S_\alpha v))_{\alpha\in\Gamma}$.

\subsection{Functional Basis}

To each vector of the basis is associated an $\alpha$ in $\Gamma$.
Take the set of functions $\theta$ of the Annexes \ref{an:Basis} and set $\iota_{\{(n,1)\}}(x,y)=\theta_{\{n\}}(x)$, $\iota_{\{(n,2)\}}(x,y)=\theta_{\{n\}}(y)$ and $\iota_{\{(n,3)\}}(x,y)=\theta_{\{n\}}(x)\theta_{\{n\}}(y)$. 

Finally for all $\alpha\in\Gamma$ set the functions $\iota_{\alpha}=\prod_{\beta\in\alpha} \iota_{\{\beta\}}$. It's easily shown that the set $(\iota_\alpha)_{\alpha\in\Gamma}$ is an orthonormal basis of $L^2([-1,1]^2,\frac{\lambda^2}{4})$.

In this basis the operator $S_z^{(n)}$ is the multiplication operator by $\theta_{\{n\}}(x)$ that is $(S_z^{(n)}f)(x,y)=\theta_{\{n\}}(x)f(x,y)$. Since in the article we deal with only with operators being sums of $S_z^{(n)}$ terms we omit the $y$ component of the functions in the article.

\section{Convergence of square summable sum of spin operators\label{convOfSum}} 

If $\sum_{n=1}^\infty |A_n|$ is convergent due to the fact that the space of bounded operators is a Banach space the infinite sum $H_=\sum_{n=1}^{\infty} A_n I_x^{(n)}$ is well defined as the limit of the partial sum $H_N=\sum_{n=1}^N A_n I_x^{(n)}$ in the topology of bounded operators. This imply furthermore that $H$ is bounded operator.

Here we will give a rigorous definition of the sum $\sum_{n=1}^\infty A_n I_x^{(n)}$ where $\sum_{n=1}^\infty |A_n|$ diverges but $\sum_{n=1}^\infty |A_i|^2$ is convergent.

As seen in the Annex \ref{an:Basis} the $I_x^{(n)}$ can be associated to the multiplication operator by the function $\theta_{\{n\}}$ which are a basis of $\mathcal{H}=L^2\left(\left[-1;1\right],\frac{\mathrm{d}x}{2}\right)$. As such we can associate $H_N$ to the multiplication operator by the function $\phi_N \defeq \sum_{n=1}^N A_n \theta_{\{n\}}$.

Remark that when $\sum_{n=1}^\infty |A_i|^2$ is convergent the infinite sum $\sum_{n=1}^\infty A_n \theta_{\{n\}}$ is well defined by the fact that the $(\theta_\alpha)_{\alpha\in\mathcal{F}}$ is a basis of $\mathcal{H}$. Now define $H$ as the multiplication operator by $\phi \defeq \sum_{n=1}^{\infty}A_n \theta_{\{n\}}$ where the convergence of the sum is understood in the $L^2\left(\left[-1;1\right],\frac{\mathrm{d}x}{2}\right)$ topology.

We know prove that $H_N$ converge to $H$ in the strong resolvent sense.

Take $D=L^{\infty}([-1,1])$ since we have a finite measure $D$ is dense in $\mathcal{H}$.

It is then a common core for $H_N$ and $H$.

Suppose $\psi\in D$:

\[
||H_N\psi-H\psi||_2\le ||\phi_N-\phi||_2 ||\psi||_\infty
\]

This show that $H_N$ converge in the strong resolvent sense to $H$.

So $H$ is a well defined unbounded operator.